\documentclass[12pt]{article}
\usepackage{amsfonts}
\usepackage{amsmath}
\usepackage{amssymb}
\usepackage{graphicx}
\usepackage{color}
\usepackage{braket}
\usepackage[utf8]{inputenc}
\usepackage{hyperref} 
\usepackage{ulem}
\pdfoutput=1

\def \be {\begin{equation}}
\def \ee {\end{equation}}
\def \ba {\begin{aligned}}
\def \ea {\end{aligned}}
\def \bea {\begin{eqnarray}}
\def \eea {\end{eqnarray}}

\begin{document}
\begin{titlepage}
\begin{flushright}
TIT/HEP- 698 \\
January, 2024
\end{flushright}
\vspace{0.5cm}
\begin{center}
{\Large \bf TBA equations and exact WKB analysis in deformed supersymmetric quantum mechanics}
\lineskip .75em
\vskip 2.5cm
{Katsushi Ito$^{a,}$\footnote{ito@th.phys.titech.ac.jp} and Hongfei Shu$^{b,c,d,}$\footnote{shuphy124@gmail.com, shu@zzu.edu.cn
}}
\vskip 2.5em
 {\normalsize\it 
$^{a}$Department of Physics, Tokyo Institute of Technology,
Tokyo, 152-8551, Japan\\
$^{b}$School of Physics and Microelectronics,
Zhengzhou University, Zhengzhou, Henan 450001, China\\
$^{c}$Beijing Institute of Mathematical Sciences and Applications, Beijing, 101408, China\\
$^{d}$Yau Mathematical Sciences Center, Tsinghua University, Beijing, 100084, China
}
\vskip 3.0em
\end{center}
  \begin{abstract}
We study the spectral problem in deformed  supersymmetric quantum mechanics with polynomial superpotential by using the exact WKB method and the TBA equations. We apply the ODE/IM correspondence to the Schr\"odinger equation with an effective potential deformed by integrating out the fermions, which admits a continuous deformation parameter. We find that the TBA equations are described by the ${\mathbb Z}_4$-extended ones. For cubic superpotential corresponding to the symmetric double-well potential, the TBA system splits into the two $D_3$-type TBA equations.  We investigate in detail this example based on the TBA equations and their analytic continuation as well as the massless limit. We find that the energy spectrum obtained from the exact quantization condition is in good agreement with the  diagonalization approach of the Hamiltonian.  
  \end{abstract}
\end{titlepage}

\tableofcontents
% \newpage

\section{Introduction}

Exact methods have particular significance in studying physics beyond perturbative approximation. The exact WKB (Wentzel-Kramers-Brillouin) analysis provides a rigorous framework to solve the bound state problem of the one-dimensional Schr\"odinger equation \cite{Balian:1978et,Voros-81,Voros-83,Silverstone-85,AKT-91,DDP-93,DDP-97,DP-99,Iwaki:2014vad}. In this method, the spectrum is solved via the exact version of the Bohr-Sommerfeld quantization condition satisfied by the exact WKB periods defined as an analytic function of the Planck constant via the Borel resummation. The Stokes phenomena of the wave function determine the global analytic structure of the WKB periods and the quantization conditions. Another exact method to solve the spectrum of the quantum integrable system is the Bethe ansatz equations which are written as the self-consistent equations for  the S-matrices of the pseudo-particles \cite{Zamolodchikov:1989cf,Klumper:1991jda,Klumper1991,Destri:1992qk,Kuniba:2010ir}. In particular, the thermodynamic Bethe ansatz (TBA) equation is a useful approach to solving the ground state energy of the two-dimensional quantum integrable field theories \cite{Yang:1968rm}. It is a non-linear integrable equation satisfied by the energies of pseudo-particles with interactions described by the S-matrices.

The correspondence between these two different approaches has been found in \cite{Dorey:1998pt,Bazhanov:1998wj}, called the ODE/IM correspondence \cite{Dorey:2007zx}.  The connection coefficients of the Schr\"odinger equation around singularities satisfy certain functional equations which appear also in the quantum integrable models. In particular, the cross-ratios of the Wronskians of the solutions with different asymptotics at infinity determine the exact WKB periods, which are shown to share the same analyticity and asymptotics with the Y-functions of the quantum integrable model.  Since it is difficult to calculate higher-order terms in the WKB series in general, the TBA equation satisfied by the Y-functions provides a convenient tool to compute the WKB periods. So far, the correspondence between the WKB periods and the Y-functions  has been studied for polynomial-type potentials \cite{Ito:2018eon,Emery:2020qqu}. See \cite{Ito:2019jio,Gabai:2021dfi,unp-zamo,Grassi:2019coc,Fioravanti:2019vxi,Fioravanti:2022bqf,Ito:2021boh,Ito:2021sjo} for various generalizations. The ODEs in the correspondence can be regarded as the quantum Seiberg-Witten (SW) curve in the Nekrasov-Shatashvili limit of the Omega background \cite{Nekrasov:2009rc}.    

To apply the one-dimensional WKB  method to a general system with many degrees of freedom, it would be good to start by considering the problem under the effective potential which introduces higher-order corrections in $\hbar$.  A centrifugal potential, which is introduced by integrating  the angular  degree of freedom,  is a typical example. For  the polynomial potential including a centrifugal term, the TBA equations deform from $A$-type to $D$-type \cite{Lukyanov:2010rn,Ito:2019jio}, where the angular momentum corresponds to the twist parameter of the boundary in the integral model.  

Another interesting example is the deformed supersymmetric quantum mechanics which we want to study in the present paper. This involves a bosonic field $x(t)$ and $N_f$ Grassmann-valued fields $\psi_i(t)$ with a superotential $W(x(t))$  \cite{Behtash:2015loa,Fujimori:2017osz, Kamata:2021jrs}\footnote{This system can also be regarded as a bosonic field coupled with the Wess-Zumino term with internal spin $S$ \cite{Stone:1988fu}. See also \cite{Balitsky:1985in,Aoyama:2001ca} for generalizations.}.
Integrating over the fermions yields a quantum system with the effective potential 
\begin{equation}\label{eq:Veff}
   V_{\rm eff}=\frac{1}{2}\big(W^{\prime }(x)\big)^2+m\hbar W^{\prime\prime}(x).
\end{equation}
 Here $m=(2k-N_f)/2$ and $k=0,\cdots,N_f$. When $N_f=1$ i.e. $m=\pm 1/2$, the potential recovers that of the original supersymmetric quantum mechanics \cite{Witten:1981nf, Witten:1982df}, with the superpotential $W(x)$. In  $SO(2n)$ gauge theories and $SU(n)$ SQCDs, there appear $\hbar$-corrections in the quantum SW curve \cite{Ito:2019twh,Ito:2020lyu}.  In this paper, we will consider the effective potential \eqref{eq:Veff} with a general value of  $m$ by applying the ODE/IM correspondence.

The resurgence structure of the deformed supersymmetric quantum mechanics has been studied via Lefschetz thimble \cite{Behtash:2015loa,Fujimori:2017osz,Behtash:2018voa}, where the complex bion is shown to be necessary for obtaining the non-perturbative structure of the model.
In \cite{Kamata:2021jrs}, the authors have studied the exact quantization condition of \cite{DDP-97} by applying the exact WKB method to investigate the resurgent structure of the ground state energy by rescaling the energy by $\hbar$. 
In this paper, we will study the spectrum of the deformed supersymmetric model in the standard WKB method and explore the correspondence to the TBA approach using the ODE/IM correspondence. The $\hbar$ deformed term $\hbar mW''(x)$ in the potential does not change the WKB analysis  but  introduces new issues to the original TBA equations. First, the rotation symmetry of the ODE changes. This requires other set  TBA equations  with different phases of $m$. Second, the deformation parameter $m$  introduces a new singularity in the TBA equations, which results in the analytic continuation of the TBA \cite{Dorey:1996re,Bazhanov:1996aq}.  Third, the critical points of the undeformed potential corresponding to the ground state energy, the higher WKB corrections for some cycles diverge \cite{Sueishi:2019xcj}. It is interesting to  consider the corresponding limit of the TBA equations. The deformation is useful to study the TBA equation around the critical point.  We will investigate these critical points for the simplest example of the models with cubic superpotential such that the symmetric double-well potential gives the undeformed potential. We can use the exact quantization condition to work out the energy spectrum numerically in detail.

%We apply the ODE/IM technique to 
We will summarize the main results in this paper.
For a deformed supersymmetric quantum mechanics with polynomial superpotential, the rotation symmetry acts as the $\mathbb{Z}_4$-symmetry on the deformation parameter. 
The Y-system then admits the $\mathbb{Z}_4$-extension. 
The deformation also changes the asymptotic behavior of the Y functions.   
We obtain $\mathbb{Z}_4$-extended TBA equations with deformed source terms.
For the cubic superpotential, the TBA equations 
%quantum-mechanical problems with polynomial potential, whose Schr\"odinger equation involves a $\hbar$-deformed term. We obtain TBA equations describing the exact WKB periods of this model, which are $\mathbb{Z}_4$ extension of the undeformed TBA equations. The undeformed quantum-mechanical problems with double-well potential corresponds to the $A_3$-type TBA equations \cite{Ito:2018eon}. Turning on the $\hbar$ deformed term in the supersymmetric quantum-mechanics, the $\mathbb{Z}_4$-extended TBA equations
reduce to two sets of $D_3$-type TBA equations \cite{Ito:2019jio}.
%, because of the remained parity symmetry in the potential. 
Using the exact quantization condition and analytic continuation of the deformation parameter, the TBA equations reproduce the energy spectrum, which is in good agreement with the ones obtained by the diagonalization approach of the Hamiltonian. The deformed TBA equations work also at the critical point of the potential although the higher-order corrections of the WKB periods diverge.

This paper is organized as follows. In section \ref{sec:exact-wkb}, we present the exact WKB analysis for the Schr\"odinger equation with $\hbar$ deformed potential and introduce the Borel resummed WKB periods. In section \ref{sec:TBA}, we apply the  ODE/IM correspondence to the Schr\"odinger equation and derive the TBA equations. A particular example with quartic potential will be presented in detail in section \ref{sec:TBA-ex}, where a simplification occurs due to the symmetries of the equations. We will numerically test our TBA equations to compare them with the WKB periods. In section \ref{sec:ana-con}, we first study the analytic continuation of TBA for the deformation parameter $m$, and then study the zero energy limit. In section \ref{sec:voros}, we study the Voros spectrum by combining our TBA equations with the exact quantization condition. Finally, we will conclude and present possible future directions in section \ref{sec:con}.

\section{Exact WKB for deformed supersymmetric quantum mechanics}\label{sec:exact-wkb}
We study the Schr\"odinger equation with an effective potential $V_{\rm eff}(x)$, where the potential contains a $\hbar$ correction:
\begin{align}\label{eq:sch-eq1}
    \left(-{\hbar^2\over2}{d^2\over dx^2}+V_{\rm eff}(x)\right)\psi(x)&=E\psi(x).
\end{align}
A typical example is  supersymmetric quantum mechanics, where the correction arises by integrating out the fermions. 
We consider more general potential which is given by $V_{\rm eff}(x)=\frac{1}{2}(W^{\prime}(x))^2+\hbar m W^{\prime\prime}(x)$. 
$W(x)$ is a superpotential, where we consider the case that $W$ is a polynomial in $x$. $m$ is a deformation parameter of the model, where $m=\pm 1/2$ corresponds to the supersymmetric quantum mechanics. We write the Schr\"odinger equation \eqref{eq:sch-eq1} in the form:
\begin{equation}\label{eq:susy-sch}
    \Big(-\hbar^2\frac{d^2}{dx^2}+Q_0(x)+\hbar Q_1(x)\Big)\psi(x)=0,
\end{equation}
where $Q_0(x)=\big(W^\prime(x)\big)^2-2E$ and $Q_1(x)=2m W^{\prime\prime}(x)$. 
Now we study the exact WKB analysis of Eq. \eqref{eq:susy-sch}.
Substituting the WKB ansatz for the wave function
\begin{equation}
    \psi(x)=\exp\Big(\frac{1}{\hbar}\int P(x^{\prime})dx^{\prime}\Big)
\end{equation}
into \eqref{eq:susy-sch}, we obtain   the Riccati equation for $P(x)$:
\begin{equation}
    \hbar\frac{d}{dx}P(x)+P(x)^{2}-Q_0(x)-\hbar Q_1(x)=0.
\end{equation}
Applying the power series expansion $ P(x)=\sum_{n=0}^\infty \hbar^n p_n(x)$ into the Riccati equation, we obtain
\begin{equation}
    \frac{d}{dx}p_{n-1}+\sum_{l=0}^{n}p_{n-l}p_{l}=Q_{0}\delta_{0,n}+Q_{1}\delta_{1,n},\quad n\geq 0,
\end{equation}
where $p_{-1}(x)=0$. 
 One can solve $p_n(x)$ recursively. The first several examples of $p_n$ are
\begin{equation}
    \begin{aligned}
        p_0&=\pm \sqrt{Q_0},\\
        p_1&=\frac{Q_{1}}{2p_{0}}-\frac{1}{2}\frac{d}{dx}\log p_{0},\\
        p_2&=-{Q_1 \over 8Q_0^{3/2}}+{Q''_0 \over 48 Q_0^{3/2}}+d(*),\\
        p_3&={Q_1^3\over 16 Q_0^{5/2}}-{Q_1 Q''_0\over 32 Q_0^{5/2}}+{Q_1''\over 48 Q_0^{3/2}}+d(*),\\
        p_4&=-{5\over128}{Q_1^4\over Q_0^{7/2}}+{5\over 128}{Q_1^2 Q''_0\over Q_0^{7/2}}
-{7\over 1536}{(Q''_0)^2\over Q_0^{7/2}}
-{1\over32}{Q_1 Q''_1\over Q_0^{5/2}}+{Q^{(4)}_0\over 768 Q_0^{5/2}}+d(*).
    \end{aligned}
\end{equation}
Here $d(*)$ denotes the total derivative term, which is irrelevant to the computation of the higher WKB periods. 
We are interested in the period integrals of $P(x)$ on the Riemann surface $\Sigma$, which is called  the WKB curve:
\begin{equation}\label{eq:WKB-curve}
    y^2=Q_0(x).
\end{equation}
Given a one-cycle $\gamma$ on $\Sigma$, one can define the WKB period
\begin{equation}\label{eq:qp1}
    \Pi_\gamma=\oint_\gamma P(x)dx=\sum_{n=0}^\infty \hbar^n\Pi_\gamma^{(n)},
\end{equation}
where 
\begin{equation}
    \Pi_\gamma^{(n)}=\oint_\gamma p_n(x)dx.
\end{equation}
Note that in contrast with the undeformed case $Q_1=0$,  we need to consider the odd terms in the expansion of the periods.

When $W(x)$ is a polynomial in $x$ with degree $N$, the WKB curve becomes a hyper-elliptic curve\footnote{Note that the WKB curve is the same as the Seiberg-Witten curve for $SU(N-1)$ gauge theory, where $E$ corresponds to the QCD scale $\Lambda^{N-1}$. } of degree $2N-2$. 
There are $2N-2$ turning points $x_1,\dots,x_{2N-2}$ which are zeros of $Q_0(x)$.
For the case where the turning points are real and distinct ($x_1<\dots <x_{2N-2}$),
one can define a set of one-cycles $\gamma_1,\dots, \gamma_{2N-1}$ on $\Sigma$, where $\gamma_{j}$ goes around the interval $[x_{2N-2-j},x_{2N-1-j}]$ $(j=1,\dots, 2N-2)$.
The cycles $\gamma_j$ for odd (even) $j$ corresponds to the allowed (forbidden) region in classical mechanics.
We choose the orientation of the cycles such that the classical WKB period is purely positive imaginary or real positive for the forbidden or allowed cycle respectively.

One can calculate the WKB periods systematically  by expressing them  in terms of  the  periods of   a basis of meromorphic differential  on  $\Sigma$ \eqref{eq:WKB-curve}
\begin{equation}\label{eq:RS-basis}
    \frac{x^a}{y}dx,\quad a=0,\cdots,2N-4.
\end{equation}
We define the periods
\begin{equation}
    \hat{\Pi}_{\gamma,a}=\oint_\gamma \frac{x^a}{y} dx.
\end{equation}
Since the differential $p_n(x)dx$ defines a meromorphic differential on  $\Sigma$, one thus can always expand it by using the basis \eqref{eq:RS-basis}
\begin{equation}
    p_n(x)dx=\sum_{a=0}^{2N-4}B_a^{(n)}\frac{x^{a}}{y}dx+d(\ast),
    \label{eq:decomp1}
\end{equation}
from which we can express the quantum corrections of the WKB period as
\begin{equation}
    \Pi_\gamma^{(n)}=\sum_{a=0}^{2N-4}B_a^{(n)}\hat{\Pi}_{\gamma,a}.
\end{equation}
This formula is convenient for calculating higher-order corrections from the classical periods.

We will illustrate the calculation of the WKB periods for a cubic superpotential $W(x)$.
For simplicity, we assume that $W(x)=x^3/3-\frac{u_2}{2} x$ so that $W'(x)=x^2-\frac{u_2}{2}$. In this example, $Q_0$  and $Q_1$  are  given by
\begin{equation}
    \begin{aligned}
        Q_0(x)=x^4-u_2 x^2+\frac{u_2^2}{4}-2E,\quad Q_1(x)=4mx.
    \end{aligned}
\end{equation}
The system corresponds to the symmetric double-well potential  deformed by $Q_1$.
Let us consider the case, where all the turning points are real and labelled by $-a<-b<b<a$ with
\begin{equation}
    a=\sqrt{\frac{u_2}{2}+\sqrt{2E}},\quad b=\sqrt{\frac{u_2}{2}-\sqrt{2E}}.
\end{equation}
The cycles $\gamma_1$, $\gamma_2$, and $\gamma_3$  are defined as the curves encircling the segments $(b,a)$, $(-b,b)$ and $(-a,-b)$.
The corresponding classical WKB periods are
\begin{equation}
    \begin{aligned}
        \Pi_{\gamma_1}^{(0)}=2\int_{b}^{a}y dx,\quad
        \Pi_{\gamma_2}^{(0)}=2\int_{b}^{-b}y dx,\quad
        \Pi_{\gamma_3}^{(0)}=2\int_{-a}^{-b}y dx\,.
    \end{aligned}
\end{equation}
Here $\Pi^{(0)}_{\gamma_1}=\Pi_{\gamma_3}^{(0)}$ is purely imaginary with  a positive imaginary part and $\Pi^{(0)}_{\gamma_2}$ is real positive.
$\gamma_1$ are $\gamma_3$ are allowed cycles and $\gamma_2$ is a forbidden cycle.
We calculate the coefficients $B_a^{(n)}$ based on the decomposition \eqref{eq:decomp1}.
We find that
\begin{align}
\Pi^{(0)}_\gamma&={2a^2b^2\over 3} \Pi_{\gamma, 0}-{a^2+b^2\over 3}\Pi_{\gamma, 2},\label{eq:DW-clas-peri}\\
\Pi^{(1)}_\gamma&=2m \Pi_{\gamma, 1},\label{eq:DW-line-corr}\\
\Pi^{(2)}_\gamma&={(a^2+b^2)(-1+12m^2)\over 6(a^2-b^2)^2}\Pi_{\gamma,0}-{a^4+b^4+2 a^2 b^2(-5+48 m^2) \over 24 a^2 b^2 (a^2-b^2)^2}\Pi_{\gamma,2}, \label{eq:DW-sec}
\\
\Pi^{(3)}_{\gamma}&=0,\label{eq:DW-thi}\\
\Pi^{(4)}_{\gamma}&=B_{0}^{(4)}\Pi_{\gamma,0}+B^{(4)}_2\Pi_{\gamma,2}, \label{eq:DW-for}
\end{align}
where
\begin{align}
B^{(4)}_0&={1\over 2880 a^4 b^4(a^2-b^2)^6}
\Bigl\{14(a^{12}+b^{12})-3 (a^{10}b^2+a^2 b^{10})(17+80 m^2) \nonumber\\
&+30(a^8 b^4+a^4 b^8)(97-1568 m^2+3456 m^4)
+10 a^6 b^6 (859-15120 m^2+28416 m^4)\Bigr\},
\\
B^{(4)}_2&=-{(a^2+b^2) \over 5760 a^6 b^6 (a^2-b^2)^6}
\Bigl\{ 56(a^{12}+b^{12})-2(a^{10}b^2+a^2 b^{10})(89+320 m^2) \nonumber\\
&+12 (a^8 b^4+a^4 b^8) (59+160 m^2+960 m^4)
+2a^6 b^6 (6671-123840 m^2+234240 m^4)
\Bigr\}.
\end{align}
Here the period integrals are evaluated in terms of the elliptic integrals:
\begin{align}
\Pi_{\gamma_1,0}&=-{2i\over a}K(k), \quad \Pi_{\gamma_2,0}={4\over a}K(k'),\\
\Pi_{\gamma_1,1}&=-i\pi , \quad \Pi_{\gamma_2,0}=0,\\
\Pi_{\gamma_1,2}&=-2ia E(k), \quad \Pi_{\gamma_2,2}=4a(K(k')-E(k')),
\end{align}
with $k=\sqrt{a^2-b^2}/a$ and $k'=b/a$. $K(k)$ ($E(k)$) is the complete elliptic integrals of the first (second) kind.
We observe that $\Pi^{(2n+1)}_{\gamma}=0$ for $n\geq 1$ so that only $\Pi^{(1)}_{\gamma_1}$ is non-zero.

The series \eqref{eq:qp1} represents an asymptotic series in $\hbar$. The Borel resummation provides a  convenient approach to evaluate the WKB series for finite $\hbar$ exactly.
We first define the Borel summation of the series \eqref{eq:qp1}
\begin{align}
    \hat{\Pi}_\gamma(\xi)&=\sum_{n=0}^\infty {\xi^n\over n!}\Pi_\gamma^{(n)},
\end{align}
which defines an analytic function in the $\xi$-plane.
By the Laplace transformation along the ray from the origin to infinity on the complex $\xi$-plane, one obtains the analytically continued form of the periods:
\begin{align}\label{eq:Borel-resum}
    s_{\varphi}[\Pi_\gamma](\hbar)&=\int_0^{\infty e^{i\varphi}}e^{-i{\xi\over \hbar}}\hat{\Pi}_\gamma(\xi).
\end{align}
The $s_{\varphi}$ shows the discontinuity due to the singularity of $\hat{\Pi}(\xi)$ in the $\xi$-plane, which is formulated as the origin of the TBA equations satisfied by  the quantum periods.
We confirmed  the singularity structure in the $\xi$-plane numerically by calculating higher-order corrections as we did for the quartic potential \cite{Ito:2018eon}.

\section{TBA equations}\label{sec:TBA}

In this section, we generalize the ODE/IM correspondence to the Schr\"odinger equation \eqref{eq:susy-sch} for the degree $N$ polynomial $W(x)$:

\begin{equation}
 W(x)=\sum_{a=0}^{N}u_{a}x^{a}.
\end{equation}
Rescaling the coordinate $x$,  the variables $u_a$ and $E$ by
\begin{equation}
    x=\hbar^{\frac{2}{2N}}z,\quad u_{a}=\hbar^{\frac{2N-2a}{2N}}b_{a},\quad E=\hbar^{2\frac{2N-2}{2N}}\hat{E},\quad W(x)=\hbar^{\frac{2N}{2N}}\hat{W}(z),\quad m=\hat{m},
\end{equation}
the Schr\"odinger equation \eqref{eq:sch-eq1} becomes
\begin{equation}\label{eq:sch-eq2}
    \Big(-\frac{d^{2}}{d z^{2}}+\big(\hat{W}^\prime(z)\big)^{2}-2\hat{E}+2\hat{m}\hat{W}^{\prime\prime}(z)\Big)\hat{\psi}(z)=0,
\end{equation}
where 
\begin{equation}
    \hat{W}(z)=\sum_{a=0}^{N}b_{a}z^{a}.
\end{equation}
Under the rotation
\begin{equation}\label{eq:rot1}
(z,b_{a},\hat{E},\hat{m})\to(\omega z,\omega^{N-a}b_{a},\omega^{2N-2}\hat{E},\omega^{N}\hat{m}),
\end{equation}
the ODE \eqref{eq:sch-eq2} becomes
\begin{equation}
    \Big(-\frac{d^{2}}{d z^{2}}+\omega^{2N}\big(\hat{W}^\prime(z)\big)^{2}-2\omega^{2N}\hat{E}+2\omega^{2N}\hat{m}\hat{W}^{\prime\prime}(z)\Big)\hat{\psi}(z)=0.
\end{equation}
Then the  ODE \eqref{eq:sch-eq2} is invariant under the rotation \eqref{eq:rot1} with  $\omega=e^{\frac{2\pi i}{2N}}$.  The fastest decaying solution of the ODE at infinity around the real positive axis behaves as
\begin{equation}
    \hat{y}(z,b_a, \hat{E}, \hat{m})\sim \frac{1}{2i}z^{n_N}\exp\Big(-\frac{z^{2N}}{N}\Big),
\end{equation} 
where $n_N=-\frac{N-1}{2}-B_{N,N}$. The coefficient $B_{N,N}$ is determined by
\begin{equation}
    \sqrt{\big(\hat{W}^\prime(z^{\prime})\big)^{2}-2\hat{E}+2\hat{m}\hat{W}^{\prime\prime}(z^{\prime})}\sim z^{\prime\frac{2N-2}{2}}\big(1+\sum_{n=1}^{\infty}B_{N,n}z^{\prime-n}\big),\quad z^\prime \to\infty.
\end{equation}
Under the rotation \eqref{eq:rot1}, one finds
\begin{equation}
    B_{N,N}(\omega z,\omega^{N-a}b_{a},\omega^{2N-2}\hat{E},\omega^{N}\hat{m})\to\omega^{N}B_{N,N}(z,b_{a},\hat{E},\hat{m}).
\end{equation}
Note that $\hat{y}$ is a subdominant solution in the sector ${\cal S}_0: |{\rm arg}(z)|<\frac{\pi}{2N}$. The subdominant solution in the sector ${\cal S}_k: |{\rm arg}(z)-\frac{2k\pi}{2N}|<\frac{\pi}{2N}$ $(k\in {\mathbb Z})$ can be obtained from $\hat{y}$ by
\begin{equation}
    \hat{y}_k(z, b_a, \hat{E}, \hat{m})=\omega^{\frac{k}{2}} \hat{y}(\omega^{-k}z, \omega^{-k(N-a)}b_a, \omega^{-k(2N-2)}\hat{E}, \omega^{-kN}\hat{m}).
\end{equation}
We then denote the Wronskian of $\hat{y}_{k_1}$ and $\hat{y}_{k_2}$ by
\begin{equation}
    \hat{W}_{k_1,k_2}(b_a,\hat{E},\hat{m}):= \hat{y}_{k_1}(z,b_a,\hat{E},\hat{m})\partial_z\hat{y}_{k_2}(z,b_a,\hat{E},\hat{m})- \hat{y}_{k_2}(z,b_a,\hat{E},\hat{m})\partial_z\hat{y}_{k_1}(z,b_a,\hat{E},\hat{m})
\end{equation}
and introduce the notation
\begin{equation}
    f^{[j]}(z,b_{a},\hat{E},\hat{m})=f(\omega^{-j/2}z,\omega^{-j(N-a)/2}b_{a},\omega^{-j(2N-2)/2}\hat{E},\omega^{-jN/2}\hat{m}).
\end{equation}
It is then easy to find
\begin{equation}
    \hat{W}_{k_{1}+1,k_{2}+1}(z,b_{a},\hat{E},\hat{m})=\hat{W}_{k_{1},k_{2}}^{[2]}(z,b_{a},\hat{E},\hat{m})
\end{equation}
and
\begin{equation}
    \hat{W}_{k,k+1}(z,b_{a},\hat{E},\hat{m})=\omega^{(-1)kB_{N,N}}.
\end{equation}

We then introduce the Y-function
\begin{equation}
    \begin{aligned}
        {\cal Y}_{2j}(b_{a},\hat{E},\hat{m})&=\frac{\hat{W}_{-j,j}\hat{W}_{-j-1,j+1}}{\hat{W}_{-j-1,-j}\hat{W}_{j,j+1}}(b_{a},\hat{E},\hat{m}),\\{\cal Y}_{2j+1}(b_{a},\hat{E},\hat{m})&=\frac{\hat{W}_{-j-1,j}\hat{W}_{-j-2,j+1}}{\hat{W}_{-j-2,-j-1}\hat{W}_{j,j+1}}(b_{a},\hat{E},\hat{m}),
    \end{aligned}
\end{equation}
which satisfies the Y-system
\begin{equation}
    {\cal Y}_{s}^{[+1]}(b_{a},\hat{E},\hat{m}){\cal Y}_{s}^{[-1]}(b_{a},\hat{E},\hat{m})=\big(1+{\cal Y}_{s-1}(b_{a},\hat{E},\hat{m})\big)\big(1+{\cal Y}_{s+1}(b_{a},\hat{E},\hat{m})\big),
\end{equation}
where $s=1, \dots, 2N-3 $.
This Y-system represents the relation among the functions  of $b_a$, $\hat{E}$, which is difficult to solve.
To obtain the TBA equations, it is convenient to go back to the Schr\"odinger equation in variable $x$. The solutions $\hat{y}$ and $\hat{y}_k$ then become
\begin{equation}
    {y}(x, u_a, E, {m}, \hbar)\equiv \hat{y}(z, b_a, \hat{E}, \hat{m}) ,\quad {y}_k(x,u_a, E, m, \hbar)\equiv \omega^{\frac{k}{2}}y(x, u_a, E, e^{-i\pi k}{m}, e^{i\pi k}\hbar).
\end{equation} 
The Wronskian of $y_{k_1}$ and $y_{k_2}$, $W_{k_1,k_2}$, is related to $\hat{W}_{k_1,k_2}$ by
\begin{equation}
    {W}_{k_{1},k_{2}}(u_{a},E,{m}):=\Big({y}_{k_{1}}\partial_{x}{y}_{k_{2}}-{y}_{k_{2}}\partial_{x}{y}_{k_{1}}\Big)=\hbar^{-\frac{1}{N}}\hat{W}_{k_{1},k_{2}}.
\end{equation}

In terms of the Wronskians, the Y-function is defined by
\begin{equation}
    \begin{aligned}
       {Y}_{2j}(\hbar,u_{a},E,{m})&=\frac{{W}_{-j,j}{W}_{-j-1,j+1}}{{W}_{-j-1,-j}{W}_{j,j+1}}(\hbar,u_{a},E,{m}),\\
        {Y}_{2j+1}(\hbar,u_{a},E,{m})&=\frac{{W}_{-j-1,j}{W}_{-j-2,j+1}}{{W}_{-j-2,-j-1}{W}_{j,j+1}}(\hbar,u_{a},E,{m}),
    \end{aligned}
\end{equation}
which satisfies the Y-system written as the functional relation for $\hbar$ and $m$ with fixed $b_a$ and $E$:
\begin{equation}
    {Y}_{s}(e^{\frac{\pi i}{2}}\hbar,e^{-\frac{\pi i}{2}}{m}){Y}_{s}(e^{-\frac{\pi i}{2}}\hbar,e^{\frac{\pi i}{2}}{m})=\big(1+{Y}_{s-1}\big)\big(1+{Y}_{s+1}\big)(\hbar,{m}), \quad s=1, \dots, 2N-3.
\end{equation}
Labelling the Y-functions with different phase factors of  $m$ by
\begin{equation}
    Y_{a,s}(\hbar)=Y_s(\hbar, e^{\frac{a\pi i}{2}}m),\quad a=0,1,2,3,
\end{equation}
we find ${\mathbb Z}_4$-extended Y-system \footnote{A similar procedure has been used in \cite{Masoero:2010is,Fioravanti:2022bqf}.}
\begin{equation}\label{eq:Y-system-Z4}
    \begin{aligned}
        &Y_{0,s}(\theta-\frac{\pi i}{2})Y_{2,s}(\theta+\frac{\pi i}{2})=\big(1+Y_{1,s-1}\big)\big(1+Y_{1,s+1}\big)(\theta),\\
        &Y_{1,s}(\theta-\frac{\pi i}{2})Y_{3,s}(\theta+\frac{\pi i}{2})=\big(1+Y_{2,s-1}\big)\big(1+Y_{2,s+1}\big)(\theta),\\
        &Y_{2,s}(\theta-\frac{\pi i}{2})Y_{0,s}(\theta+\frac{\pi i}{2})=\big(1+Y_{3,s-1}\big)\big(1+Y_{3,s+1}\big)(\theta),\\
        &Y_{3,s}(\theta-\frac{\pi i}{2})Y_{1,s}(\theta+\frac{\pi i}{2})=\big(1+Y_{0,s-1}\big)\big(1+Y_{0,s+1}\big)(\theta)
    \end{aligned}
\end{equation}
with $s=1, \dots, 2N-3$. Here we have introduced the spectral parameter $\theta$ by $\hbar=e^{-\theta}$. We now derive the TBA equations associated with the Y-system \eqref{eq:Y-system-Z4}.
We assume that $Y_{a,s}(\theta)$ are analytic in the strip $|{\rm Im}\theta|\leq {\pi\over 2}$, which can be checked in the following numerical analysis. 
For  $\hbar\rightarrow 0$ ($\theta\rightarrow \infty$),
the asymptotic behavior of Y-functions is given by
\begin{equation}
    \begin{aligned}
        \log Y_{a,2k+1}(\theta)\sim&-\frac{1}{i\hbar}\oint_{\gamma_{2k+1}}p_{0}dx -i^a\oint_{\gamma_{2k+1}}p_{1}dx+{\cal O}(\hbar)=:-\frac{m_{2k+1}}{\hbar}+{m}_{a,2k+1}^{(\frac{1}{2})}+{\cal O}(\hbar)\\
        \log Y_{a,2k}(\theta)\sim&-\frac{1}{\hbar}\oint_{\gamma_{2k}}p_{0}dx-i^a\oint_{\gamma_{2k}}p_{1}dx+{\cal O}(\hbar)=:-\frac{m_{2k}} {\hbar}+m_{a,2k}^{(\frac{1}{2})}+{\cal O}(\hbar),\quad \theta\to \infty.
    \end{aligned}
\end{equation}
where the contours $\gamma_{2k}$ and $\gamma_{2k+1}$ are defined in the previous section. The masses $m_s$ and their corrections $m^{({1\over2})}_{a,s}$ at order $\hbar=(\hbar^2)^{1\over2}$, which we will call the linear correction below,  are given by
\begin{equation}
    \begin{aligned}
        m_{2k+1}&=\frac{1}{i}\oint_{\gamma_{2k+1}}p_0 dx,\quad m_{2k}=\oint_{\gamma_{2k}}p_0 dx \\
        m_{a,s}^{(\frac{1}{2})}&=-i^a\oint_{\gamma_s}p_1 dx,\qquad\quad s=2k, 2k+1. 
    \end{aligned}
\end{equation}
Here we have defined $m_s$ such that they are positive when all the turning points, zeros of $Q_0(z)$, are real and different.

From the analyticity and the asymptotics, one converts the Y-system \eqref{eq:Y-system-Z4} to a set of the $\mathbb{Z}_4$ extended TBA equations
\begin{equation}\label{eq:TBA}
    \begin{aligned}
    \log Y_{0,s}=&-m_{s}e^{\theta}+m_{0,s}^{(\frac{1}{2})}+K_{+}\star\Big[\log\big(1+Y_{1,s-1}\big)+\log\big(1+Y_{1,s+1}\big)\Big]\\
    &+K_{-}\star\Big[\log\big(1+Y_{3,s-1}\big)+\log\big(1+Y_{3,s+1}\big)\Big],\\
    \log Y_{1,s}=&-m_{s}e^{\theta}+m_{1,s}^{(\frac{1}{2})}+K_{+}\star\Big[\log\big(1+Y_{2,s-1}\big)+\log\big(1+Y_{2,s+1}\big)\Big]\\
    &+K_{-}\star\Big[\log\big(1+Y_{0,s-1}\big)+\log\big(1+Y_{0,s+1}\big)\Big],\\
    \log Y_{2,s}=&-m_{s}e^{\theta}-m_{0,s}^{(\frac{1}{2})}+K_{+}\star\Big[\log\big(1+Y_{3,s-1}\big)+\log\big(1+Y_{3,s+1}\big)\Big]\\
    &+K_{-}\star\Big[\log\big(1+Y_{1,s-1}\big)+\log\big(1+Y_{1,s+1}\big)\Big],\\
    \log Y_{3,s}=&-m_{s}e^{\theta}-m_{1,s}^{(\frac{1}{2})}+K_{+}\star\Big[\log\big(1+Y_{0,s-1}\big)+\log\big(1+Y_{0,s+1}\big)\Big]\\
    &+K_{-}\star\Big[\log\big(1+Y_{2,s-1}\big)+\log\big(1+Y_{2,s+1}\big)\Big],  \qquad s=1, \dots, 2N-3,
    \end{aligned}
\end{equation}
where $\star$ denotes the convolution $(f\star g)(\theta)=\int_{-\infty}^{\infty}f(\theta-\theta')g(\theta')d\theta'$.  The kernel functions $K_{\pm}(\theta)$ are defined by
\begin{equation}
    \begin{aligned}
    K_{\pm}(\theta)=\frac{1}{4\pi}\Big(\frac{1}{\cosh\theta}\pm i\frac{\sinh\theta}{\cosh\theta}\Big).
    \end{aligned}
\end{equation}
More compactly, we can write the TBA equations as
\begin{equation}
    \begin{aligned}
        \log Y_{a,s}=&-m_{s}e^{\theta}+m_{a,s}^{(\frac{1}{2})}+K_{+}\star L_{a+1,s-1}+K_{+}\star L_{a+1,s+1}\\&+K_{+}\star L_{a+3,s-1}+K_{+}\star L_{a+3,s+1},\quad a \equiv a+4,
    \end{aligned}
\end{equation}
where $L_{a,s}(\theta)=\log\big(1+Y_{a,s}(\theta)\big)$.

Note that the $\tanh(\theta)$ part in the kernels does not decay at large $\theta$, which causes difficulty in the numerical study \footnote{A similar kernel also appears in the TBA equations for the minimal surface with light-like boundary in AdS$_5$ spacetime \cite{Alday:2010vh}.}.
This $\tanh(\theta)$ term also leads to odd power terms in the $\hbar$ expansion of the WKB period. The linear correction $m_{a,s}^{(\frac{1}{2})}$ plays the role of chemical potential in the TBA equations, which will also change  the asymptotic at $\theta\to -\infty$. When $m\to 0$, all the linear corrections will vanish, $m_{a,s}^{(\frac{1}{2})}\to 0$, the ${\mathbb Z}_4$ TBA equations will reduce to the  $A$-type TBA as in \cite{Ito:2018eon}.

\paragraph{Wall-crossing of the TBA equations}  We  have focused on the case where all the turning points are along the real axis, such that all the masses are positive and real. Moving the turning points away from the real axis, the masses become complex values generically. In this case, one shifts the spectral parameter $\theta$ to keep the leading order $m_s/\hbar$ being positive and real. Then the TBA equations can be derived in a similar way
\begin{equation}
    \begin{aligned}
        \log Y_{a,s}(\theta-i\phi_{s})=&-|m_{s}|e^{\theta}+m_{a,s}^{(0)}+K_{+;s,s-1}\star\widetilde{L}_{a+1,s-1}+K_{+;s,s+1}\star\widetilde{L}_{a+1,s+1}\\&+K_{-;s,s-1}\star\widetilde{L}_{a+3,s-1}+K_{-;s,s+1}\star\widetilde{L}_{a+3,s+1},
    \end{aligned}
\end{equation}
where $\phi_s={\rm Arg}(m_s)$, the kernel $K_{\pm;r,s}$ and $\widetilde{L}_{a,s}$ are 
\begin{equation}
    K_{\pm;r,s}=K_{\pm}(\theta-i\phi_{r}+i\phi_{s}),\quad \widetilde{L}_{a,s}=\log\big(1+Y_{a,s}(\theta-i\phi_{s})\big).
\end{equation}
It is worth noting that poles appear in the kernel $K_{\pm;s,s\pm 1}$ when
\begin{equation}
    |\phi_s-\phi_{s\pm 1}|<\frac{\pi}{2},
\end{equation}
which constrains the moduli parameters of $Q_0$.
This region is called the minimal chamber in the moduli space of the WKB curve. As $\phi_{s}-\phi_{s\pm 1}$ crosses ${\pi\over2}$, we should modify the TBA equations to include the residue of the pole in the kernel, which is known as the wall-crossing of the TBA equations \cite{Gaiotto:2009hg,Alday:2010vh,toledo-thesis,toledo}. Since the phase differences $\phi_{s}-\phi_{s\pm 1}$ are independent of the deformation term $\hbar Q_1$ of the potential, the wall-crossing of our TBA equations can be performed in a similar way as the one in \cite{Ito:2018eon}.

\paragraph{Analytic continuation of the TBA equations} In contrast to the wall-crossing phenomenon, when $m$ changes, singularities also appear in the function  $L_{a,s}$ in the TBA equations. 
Crossing these singularities, one should also pick up their contribution \cite{Dorey:1996re}. We will call this phenomenon the analytic continuation of the TBA equations.

\section{TBA equations for deformed double-well potential} \label{sec:TBA-ex}

In this section, we consider the TBA equations corresponding to the superpotential $W(x)=\frac{1}{3}x^3-\frac{1}{4}x$, where the Schr\"odinger equation \eqref{eq:sch-eq1} becomes
\begin{equation}
    \Big(-\hbar^{2}\frac{d^{2}}{dx^{2}}+\big(x^{2}-\frac{1}{4}\big)^{2}-2E+4\hbar mx\Big)\psi(x)=0.
\end{equation}
For the undeformed case $m=0$, this is an example with symmetric double-well potential which is well studied by using the exact WKB method \cite{Zinn-Justin:1982hva,DDP-97,Alvarez04,Zinn-Justin:2004vcw,Zinn-Justin:2004qzw,Jentschura:2004jg,Dunne:2014bca} as well as the TBA equation \cite{Ito:2018eon}.
For the deformed case, $m\neq 0$, including the supersymmetric point at $m=1/2$, has been studied by using the WKB approach \cite{Sueishi:2019xcj,Kamata:2021jrs}.

The TBA equations \eqref{eq:TBA} includes $12$ equations of $Y_{a,s}$, $a=1,2,3,4$ and $s=1,2,3$. We will focus on the region $0<E<\frac{1}{32}$, where the turning points are all real. The classical mass $m_s$ and the linear correction $m_{a,s}^{(\frac{1}{2})}$ can be evaluated by using \eqref{eq:DW-clas-peri} and \eqref{eq:DW-line-corr}, which lead to
\begin{equation}
    m_1=m_3,\quad m_{a,2}^{(\frac{1}{2})}=0,\quad m_{a,1}^{(\frac{1}{2})}=-m_{a,3}^{(\frac{1}{2})}.
\end{equation}
From the TBA equations \eqref{eq:TBA}, one thus finds
the following identification of the Y-functions
\begin{equation}
    \begin{aligned}
        Y_{0,1}&=Y_{2,3},\quad Y_{1,1}=Y_{3,3},\quad Y_{2,1}=Y_{0,3},\\\quad Y_{3,1}&=Y_{1,3},\quad        Y_{1,2}=Y_{3,2},\quad Y_{0,2}=Y_{2,2}.
    \end{aligned}
\end{equation}
The $\mathbb{Z}_4$-extended TBA equations decouple into two sets of equations:
\begin{equation}\label{eq:TBA1}
    \begin{aligned}
       \log Y_{0,1}=&-m_{1}e^{\theta}+2\pi i m+K\star L_{1,2},\\
       \log Y_{2,1}=&-m_{1}e^{\theta}-2\pi i m+K\star L_{1,2},\\
       \log Y_{1,2}=&-m_{2}e^{\theta}+K\star\Big[L_{0,1}+L_{2,1}\Big]
    \end{aligned}
\end{equation}
and
\begin{equation}\label{eq:TBA2}
    \begin{aligned}
        \log Y_{1,1}=&-m_{1}e^{\theta}-2\pi  m+K\star L_{0,2},\\
        \log Y_{3,1}=&-m_{1}e^{\theta}+2\pi  m+K\star L_{0,2},\\
        \log Y_{0,2}=&-m_{2}e^{\theta}+K\star\Big[L_{1,1}+L_{3,1}\Big],
    \end{aligned}
\end{equation}
where the kernel $K$ is defined by $K=K_++K_-=\frac{1}{2\pi \cosh\theta}$. It is also useful to introduce the S-matrix  associated with the ${\mathbb Z}_4$-model \cite{Koberle:1979sg}:
\begin{equation}\label{eq:S-matrix}
    S(\theta)={e^{\theta}-i\over e^{\theta}+i},
\end{equation}
where $K(\theta)={1\over 2\pi i}{d\over d\theta}\log S(\theta)$. It is useful to introduce the notation
\begin{equation}
    \hat{Y}_{0,1}=e^{-2\pi i m}Y_{0,1}=e^{2\pi i m}Y_{2,1},\quad  \hat{Y}_{1,1}=e^{2\pi  m}Y_{1,1}=e^{-2\pi m}Y_{3,1},
\end{equation}
such that the two decoupled TBA systems can be written as
\begin{equation}\label{eq:new-TBA1}
    \begin{aligned}
        \log\hat{Y}_{0,1}=&-m_{1}e^{\theta}+K\star \log\big(1+Y_{1,2}\big),\\
        \log{Y}_{1,2}=&-m_{2}e^{\theta}+K\star \log\big(1+e^{2\pi im}\hat{Y}_{0,1}\big)+\log\big(1+e^{-2\pi im}\hat{Y}_{0,1}\big)
    \end{aligned}
\end{equation}
and
\begin{equation}\label{eq:new-TBA2}
    \begin{aligned}
        \log\hat{Y}_{1,1}=&-m_{1}e^{\theta}+K\star \log\big(1+Y_{0,2}\big),\\
        \log{Y}_{0,2}=&-m_{2}e^{\theta}+K\star \log\big(1+e^{-2\pi m}\hat{Y}_{1,1}\big)+\log\big(1+e^{2\pi m}\hat{Y}_{1,1}\big).
    \end{aligned}
\end{equation}
For later convenience, we also introduce $\hat{L}_{0,1}$ and $\hat{L}_{1,1}$ 
\begin{equation}\label{eq:Lhat}
    \begin{aligned}
        \hat{L}_{0,1}&=\log\big(1+e^{2\pi im}\hat{Y}_{0,1}\big)+\log\big(1+e^{-2\pi im}\hat{Y}_{0,1}\big),\\
        \hat{L}_{1,1}&=\log\big(1+e^{-2\pi m}\hat{Y}_{1,1}\big)+\log\big(1+e^{2\pi m}\hat{Y}_{1,1}\big).
    \end{aligned}
\end{equation}
It is worth noting that the first set of TBA equations coincides with the symmetric $D_3$ TBA-system in \cite{Ito:2019jio} with the identifications $2\pi i(l+\frac{1}{2})=m_{0,1}^{(\frac{1}{2})}=2\pi im$ and
\begin{equation}
    \hat{Y}\leftrightarrow\hat{Y}_{0,1},\quad Y_{1}\leftrightarrow{Y}_{1,2}.
\end{equation}
Let us now comment on how the TBA equations are deformed due to the $\hbar$ deformation of potential in the Schr\"odinger equation.
The Schr\"odinger equation with undeformed double-well potential corresponds to an $A_3$-type TBA system \cite{Ito:2018eon}. Adding a $\hbar$ deformed term to the potential, the TBA system is extended by ${\mathbb Z}_4$. Thanks to the symmetries of Y-functions, the $\mathbb{Z}_4$-extended TBA system finally decouples to two $D_3$-type TBA systems. 

Given the TBA equations, we can extract the asymptotics of the Y-function for $\theta\to -\infty$ and $\theta\to \infty$. As $\theta\to -\infty$, Y-functions approached constant values $Y^\ast$. Using the formula $\int_{\mathbb{R}}K(\theta) d\theta=1/2$, the TBA equations lead to the constraints on the constant values
\begin{equation}
    \begin{aligned}
        2\log\hat{Y}_{0,1}^{\ast}=&\log\big(1+{Y}_{1,2}^{\ast}\big),\\
        2\log{Y}_{1,2}^{\ast}=&\log\big(1+e^{2\pi im}\hat{Y}_{0,1}^{\ast}\big)+\log\big(1+e^{-2\pi im}\hat{Y}_{0,1}^{\ast}\big).
    \end{aligned}
\end{equation}
These equations can be solved and the constants are found to be 
\begin{equation}\label{eq:const-sol}
    \hat{Y}_{0,1}^{\ast}=2\cos(\frac{2\pi m}{3}),\quad Y_{1,2}^{\ast}=\frac{\sin(2\pi m)}{\sin(\frac{2\pi m}{3})}.
\end{equation}
The asymptotics of $Y_{a,s}(\theta)$ in the limit $\theta\rightarrow -\infty$ is determined by \eqref{eq:const-sol}. At $\theta\rightarrow \infty$, $Y_{a,s}$ behaves as $Y_{a,s}(\theta)\sim -m_{s} e^{\theta}$.

We now compare the asymptotics of $Y_{a,s}(\theta)$ with the expansion of WKB periods at large $\theta$. At $\theta\to \infty$, we can expand the kernel of the TBA equation as
\begin{equation}\label{eq:asympt1}
    -\log Y_{a,s}(\theta)\sim m_{s}e^{\theta}- m_{a,s}^{(\frac{1}{2})}+\sum_{n=1}^{\infty}m_{a,s}^{(n)}e^{(1-2n)\theta},
\end{equation}
where
\begin{equation}
    m_{a,s}^{(n)}=\frac{(-1)^{n}}{\pi}\int_{-\infty}^{\infty}d\theta e^{(2n-1)\theta}F_{a,s},\quad n\geq 1.
    \end{equation}
    $F_{a,s}$ are defined by
    \begin{equation}
    \begin{aligned}
    F_{0,1}&=F_{2,1}=L_{1,2},\quad F_{1,2}=L_{0,1}+L_{2,1},\\
    F_{1,1}&=F_{3,1}=L_{0,2},\quad F_{0,2}=L_{1,1}+L_{3,1}.
    \end{aligned}
\end{equation}
On the other hand, the quantum correction of WKB periods can be computed as discussed in Sect.2, see eq.\eqref{eq:DW-sec}-\eqref{eq:DW-for}. In Table \ref{tab:tba_wkbm05}, we compare the coefficients $m_{a,s}^{(n)}$ with the quantum correction $\Pi^{(2n)}_{\gamma_s}$ at the supersymmetric point $m=1/2$.
\begin{table}[]
    \centering
    \begin{tabular}{|c|c|c|c|c|}\hline
    $n$&    $m^{(n)}_{0,1}$ & $\Pi^{(n)}_{\gamma_1}/i$
    & $m_{1,2}^{(n)}$ & $\Pi_{\gamma_2}^{(n)}$\\\hline
   0&0.103932897990&0.103932897990&0.146983313914&0.146983313914\\1&-0.595015127256&0.595015127355&10.917186649450 &10.917186650161\\2&1.135557990286\ensuremath{\cdot10^{2}}&1.135557990512\ensuremath{\cdot10^{2}}&-1.258067561652\ensuremath{\cdot10^{3}}&-1.258067561659\ensuremath{\cdot10^{3}}\\3&-7.852584078608\ensuremath{\cdot10^{4}}&7.852584080303\ensuremath{\cdot10^{4}}&1.299686482674\ensuremath{\cdot10^{6}}&1.299686482683\ensuremath{\cdot10^{6}}\\\hline
    \end{tabular}    \caption{
 Comparison of the coefficients in the expansion based on the TBA equation and the WKB periods at $m=1/2$ with $u_2=1/2$ and $E=1/64$. The TBA equations are solved numerically by using the discretized Fourier transformation with $2^{20}$ points and the cutoff  $(-50,50)$.}
    \label{tab:tba_wkbm05}
\end{table}

The numerical calculation of the expansions results in the formula
\begin{equation}
    \begin{aligned}
        m_{0,1}^{(n)}=(-1)^{n}{\frac{1}{i}}\Pi_{\gamma_{1}}^{(2n)}(m),\quad m_{1,2}^{(n)}=\Pi_{\gamma_{2}}^{(2n)}(m).
    \end{aligned}
\end{equation}
Moreover, we have observed the discontinuity formulas of the Borel resummed WKB period numerically from the Borel-Pad\'e approximation in Sect. \ref{sec:exact-wkb}. The Y-functions are also shown to have the same discontinuity structure as in \cite{Ito:2018eon}. We thus find the identifications between Y-functions and WKB periods:
\begin{equation}
    -\log Y_{0,1}(\theta+\frac{\pi i}{2}\pm i0)=\frac{1}{\hbar}s_{\pm 0}\big(\Pi_{\gamma_{1}}\big)(\hbar),\quad-\log Y_{1,2}(\theta)=\frac{1}{\hbar}s_0(\Pi_{\gamma_{2}})(\hbar),
\end{equation}
where the Borel resummation is defined in \eqref{eq:Borel-resum}.

Similarly, we can also compare the asymptotic expansion of $-\log Y_{1,1}(\theta)$ and the WKB periods, which leads to
\begin{equation}
    m_{1,1}^{(n)}=(-1)^n\frac{1}{i}\Pi_{\gamma_1}^{(2n)}\Big|_{m\to im},\quad m_{0,2}^{(n)}=\Pi_{\gamma_2}^{(2n)}\Big|_{m\to im}.
\end{equation}
We thus can identify
\begin{equation}
    -\log Y_{1,1}(\theta+\frac{\pi i}{2}\pm i0)=\frac{1}{\hbar}s_{\pm0}\big(\Pi_{\gamma_{1}}\big)(\hbar)\Big|_{m\to im},\quad-\log Y_{0,2}(\theta)=\frac{1}{\hbar}s(\Pi_{\gamma_{2}})(\hbar)\Big|_{m\to im}.
\end{equation}

\subsection{Effective central charge}
Our TBA equations whose mass term is given by the energy of a massless particle appear as the kink limit of the TBA equations of the massive integrable quantum field theory (IQFT). The kink TBA system is characterized by the effective central charge, which describes the underlying CFT as the UV limit of the IQFT. For the TBA equations \eqref{eq:TBA1} and \eqref{eq:TBA2}, the effective central charge is defined by
\begin{equation}
    c_{\rm eff}=\hat{c}_{\rm eff}+\tilde{c}_{\rm eff},
\end{equation}
where $\hat{c}_{\rm eff}$ and $\tilde{c}_{\rm eff}$ are the ones of the decoupled system \eqref{eq:TBA1} and \eqref{eq:TBA2} respectively defined by
\begin{equation}\label{eq:ceff1}
    \begin{aligned}
        \hat{c}_{{\rm eff}}(m)=&\frac{12}{\pi^{2}}\int d\theta\Big(m_{1}e^{\theta}(L_{0,1}+L_{2,1})+m_{2}e^{\theta}L_{1,2}\Big),\quad -\frac{1}{2}<m<\frac{1}{2}
        \end{aligned}
\end{equation}
and 
\begin{equation}
    \tilde{c}_{\rm eff}(m)=\frac{12}{\pi^{2}}\int d\theta\Big(m_{1}e^{\theta}(L_{1,1}+L_{3,1})+m_{2}e^{\theta}L_{0,2}\Big),\quad -\frac{1}{2}<m<\frac{1}{2}.
\end{equation}
The integrals of the right-hand side of \eqref{eq:ceff1} can be evaluated exactly by using the boundary values \eqref{eq:const-sol} of the Y-function at $\theta\to -\infty$. Then $\hat{c}_{{\rm eff}}(m)$ is given by
\begin{equation}\label{eq:ceff-hat}
    \begin{aligned}
        \hat{c}_{{\rm eff}}(m)=&\frac{12}{\pi^{2}}\Big({\cal L}_{1}\big(\frac{1}{1+Y_{1,2}^{\ast-1}}\big)+{\cal L}_{e^{2\pi im}}\big(\frac{e^{-2\pi im}}{1+Y_{0,1}^{\ast-1}}\big)+{\cal L}_{e^{-2\pi im}}\big(\frac{e^{2\pi im}}{1+Y_{2,1}^{\ast-1}}\big)\Big) \\
        =&4(1-8m^{2}),
    \end{aligned}
\end{equation}
where ${\cal L}_c(x)$ $(c\in {\mathbb C})$ is the Rogers dilogarithm function: 
\begin{equation}
    \begin{aligned}
        {\cal L}_{c}(x)=-\frac{1}{2}\int_{0}^{x}dy\Big(\frac{c\log y}{1-cy}+\frac{\log(1-cy)}{y}\Big)=\frac{1}{2}\big(\log(x)\log(1-cx)+2{\rm Li}_{2}(cx)\big).
    \end{aligned}
\end{equation}
Here ${\rm Li}_2(x)$ is the dilogarithm.
Similarly, the effective central charge associated with the second set  of the TBA system is given by
\begin{equation}
    \begin{aligned}
        \tilde{c}_{\rm eff}(m)
        =4(1+8m^{2}).
    \end{aligned}
\end{equation}
The total effective central charge is always a constant
\begin{equation}\label{eq:ceff-full}
    c_{\rm eff}=8.
\end{equation}

In terms of masses $m_s$ and its quantum correction $m^{(1)}_s$ of order $e^{-2\theta}$, $\hat{c}_{\rm eff}$ can be expressed as
\begin{equation}\label{eq:pnp1}
    \hat{c}_{\rm eff}=-\frac{12}{\pi}\Big(m_{1}m_{2}^{(1)}+m_{2}m_{1}^{(1)}\Big),
\end{equation}
which provides a direct relation between the CFT and the WKB approaches.
The relation \eqref{eq:pnp1} between the classical WKB periods and their quantum corrections is called the perturbative/non-perturbative (PNP) relation or the quantum version of the Matone relations of the WKB periods \cite{Matone:1995rx, Sonnenschein:1995hv,Eguchi:1995jh, Dunne:2014bca,Gorsky:2014lia,Basar:2015xna,Codesido:2017dns,Basar:2017hpr}.
One can also show that the PNP relation \eqref{eq:pnp1} from the relation
\begin{align}\label{eq:PNP-Pi}
    \Pi^{(0)}_{\gamma_1}\Pi^{(2)}_{\gamma_2}-\Pi^{(0)}_{\gamma_2}\Pi^{(2)}_{\gamma_1}&=-{\pi i\over 3}
    (1-8m^2),
\end{align}
which can be proved by using Legendre's relation for the elliptic integrals.

\section{TBA equations for $m>1/2$}\label{sec:ana-con}

The TBA equations \eqref{eq:TBA1} for the deformed supersymmetric quantum mechanics are valid in the region $|m|<{1\over2}$. 
This section will study the TBA equations in the region $|m|>{1\over2}$. 
For $|m|<{1\over2}$, $Y_{0,1}$ and $Y_{1,2}$ obeying the asymptotic conditions \eqref{eq:asympt1} at $\theta\rightarrow\infty$
\begin{align}
    Y_{0,1}(\theta)&\sim m_1 e^\theta, \quad
    Y_{1,2}(\theta)\sim m_2 e^\theta,
\end{align}
and \eqref{eq:const-sol} at $\theta\rightarrow -\infty$
\begin{align}
    Y_{0,1}(\theta)&\rightarrow Y^*_{0,1}, \quad
    Y_{1,2}(\theta)\rightarrow Y^*_{1,2},
\end{align}
are entire function in the strip $|{\rm Im}\theta |\leq {\pi\over2}$. When $|m|<{1\over2}$, ${Y}_{1,2}^\ast$ is positive on the real axis. Then the Y-function $Y_{1,2}(\theta)$ does not have zeros for finite $\theta$. When $|m|={1\over2}$, constant value ${Y}_{1,2}^\ast\big|_{m=1/2}=0$. In other words, $Y_{1,2}(\theta)$ has a zero at $-\infty$. As $m$ crosses $1/2$ slightly, $Y_{1,2}^\ast$ becomes negative, which implies a zero of $Y_{1,2}(\theta)$ at finite $\theta$.
For $|m|>{1\over2}$,  $Y_{0,1}(\theta)$ and $Y_{1,2}(\theta)$ can have zero in the strip in general. Those zeros imply singularities at the left-hand side of the TBA equation. In the meantime, the function $\log\big(1+Y\big)$ in the convolution of the right-hand side of TBA will also become singular.  Therefore, the TBA equations should be modified. 
When $Y_{0,1}(\theta)$ and $Y_{1,2}(\theta)$ have $K$ zeros $\theta_{1,k}\pm i{\pi\over2}$ and $J$ zeros $\theta_{2,j}\pm i{\pi\over2}$ in the strip.
The TBA equations are modified to \cite{Dorey:1996re,Bazhanov:1996aq}\footnote{See also \cite{Fendley:1997ys,Gabai:2021dfi,Ito:2023cyz} for related developments.}
\begin{align}
    \log \hat{Y}_{0,1}(\theta)&=-m_1 e^\theta+K\star L_{1,2}-\sum_{j=1}^{J}\log S(\theta-\theta_{2,j}), \label{eq:Ana-TBA1}\\
    \log Y_{1,2}(\theta)&=-m_2 e^\theta+K\star \hat{L}_{0,1}
    -\sum_{k=1}^{K}\log S(\theta-\theta_{1,k}).\label{eq:Ana-TBA2}
\end{align}
Here $S(\theta)$ is the $S$-matrix for the kernel defined in \eqref{eq:S-matrix}, which has zero or pole at $\theta=\pm i{\pi\over2}$.
The zeros also satisfy the consistency conditions
\begin{align}
    Y_{1,2}(\theta_{2,j})&=-1, \quad Y_{0,1}(\theta_{1,k})=e^{2\pi im} \ \mbox{or} \ e^{-2\pi im}.
\end{align}
Now $\theta_{1,k}$ and $\theta_{2,j}$ satisfy
\begin{align}
\pm 2\pi im-(2N_k+1)\pi i &=-m_1 e^{\theta_{1,k}}+K\star L_{1,2}(\theta_{1,k})-\sum_{j=1}^{J}\log S(\theta_{1,k}-\theta_{2,j}),\\
-(2M_j+1)\pi i&=-m_2 e^{\theta_{2,j}}+K\star \hat{L}_{0,1}(\theta_{2,j})-\sum_{k=1}^{K}\log S(\theta_{2,j}-\theta_{1,k}).
\end{align}
where $M_j$ and $N_k$ are integers. Picking up the residues of the poles in $L$, the effective central charge \ref{eq:ceff1} is modified to
 \begin{equation}
   \begin{aligned}
       \hat{c}_{\rm eff}(m)=&\frac{12}{\pi^{2}}\int d\theta\Big(m_{1}e^{\theta}(L_{0,1}+L_{2,1})+m_{2}e^{\theta}L_{1,2}\Big)+i\frac{24}{\pi}\Big(\sum_{k=1}^{K}m_{1}e^{\theta_{1,k}}+\sum_{j=1}^{J}m_{2}e^{\theta_{2,j}}\Big),
   \end{aligned} 
\end{equation}
which expresses the excited state energy of the quantum integral model. Given TBA equations \eqref{eq:Ana-TBA1} and \eqref{eq:Ana-TBA2}, we can evaluate the integral by using the Rogers dilogarithm function
\begin{equation}
    \begin{aligned}
    &\frac{12}{\pi^{2}}\int d\theta\Big(m_{1}e^{\theta}(L_{0,1}+L_{2,1})+m_{2}e^{\theta}L_{1,2}\Big)\\
    =&\frac{12}{\pi^{2}}\Big({\cal L}_{1}\big(\frac{1}{1+Y_{1,2}^{\ast-1}}\big)+{\cal L}_{e^{2\pi im}}\big(\frac{e^{-2\pi im}}{1+Y_{0,1}^{\ast-1}}\big)+{\cal L}_{e^{-2\pi im}}\big(\frac{e^{2\pi im}}{1+Y_{2,1}^{\ast-1}}\big)\Big)\\
    &+\frac{6}{\pi^{2}}\int d\theta\Big(-4\pi i\sum_{j=1}^{J}K(\theta-\theta_{2,j})\big(L_{0,1}(\theta)+L_{2,1}(\theta)\big)-4\pi i\sum_{k=1}^{K}K(\theta-\theta_{1,k})L_{1,2}(\theta)\Big)\\
    &+\sum_{k=1}^{K}\log(-1)L_{1,2}(-\infty)+\sum_{j=1}^{J}\log(-1)\hat{L}_{0,1}\big(L_{0,1}(-\infty)+L_{2,1}(-\infty)\big),
    \end{aligned}
\end{equation}
which leads to
\begin{equation}\label{eq:ceff-Ana1}
   \begin{aligned}
       \hat{c}_{\rm eff}(m)=4(1-8m^{2}).
   \end{aligned} 
\end{equation}
For real $m$, no pole appears in the $L$-function of TBA equations \eqref{eq:new-TBA2}, the TBA equations and the effective central charge $\tilde{c}_{\rm eff}$ will thus keep the same form as the ones in $-1/2<m<1/2$. Since \eqref{eq:ceff-Ana1} has the same value as \eqref{eq:ceff-hat}, the total effective central charge will also be invariant, \eqref{eq:ceff-full}, for any $m$.   

In the next subsection, we will illustrate  a detailed analysis for the regions $1/2<m<3/4$ and $3/4<m<1$.
For $|m|<{1\over2}$, $Y_{0,1}(\theta)$ and $Y_{1,2}(\theta)$ are positive on the real $\theta$. Then they do not have zeros on the real $\theta$-axis.
When $m$ goes over ${1\over2}$, $Y^{*}_{1,2}$ becomes negative. Then $Y_{1,2}(\theta)$ has a zero $\theta_{1,1}-i{\pi\over2}$ on the real axis. $\theta_{1,k}$ start with $-\infty$ at $m={1\over2}$ and increases when $m$ becomes large. We use the TBA equation for $K=1$ and $J=0$.
When $m$ goes over ${3\over4}$, the asymptotic value of $Y_{0,1}$ changes from positive to minus.
Then $Y_{0,1}(\theta)$ has a zero at $\theta_{2,1}- i{\pi \over2}$. This is the TBA equations for $J=K=1$.
$\theta_{2,1}$ also starts from $-\infty$ to $\infty$ when $m$ increases from the value of ${3\over4}$.

\subsection{TBA for $1/2<m<1$}
We will investigate the analytic continuation of the TBA equation in the region $|m|<1$.
For $1/2<m<3/4$, there is a singularity $\theta_{1,1}+i{\pi\over2}$ on the real axis, where $\log\big(1+ Y_{0,1}(\theta)\big)$ becomes singular.
We refer to the zero of $ 1+e^{m_{0,1}}\hat{Y}_{0,1}$ as $\theta_{1,1}$, which can be obtained by solving the equation
\begin{equation}
     \log(-e^{-m_{0,1}})=i\pi-m_{0,1}=-m_{1}e^{\theta_{0,1}}+\int\frac{d\theta}{2\pi}\frac{\log\big(1+{Y}_{1,2}(\theta)\big)}{\cosh(\theta_{1,1}-\theta)}.
\end{equation}
We thus obtain the new TBA equations
\begin{equation}\label{eq:ac-TBA}
    \begin{aligned}
        \log\hat{Y}_{0,1}=&-m_{1}e^{\theta}+K\star L_{1,2},\\
        \log{Y}_{1,2}=&-m_{2}e^{\theta}+K\star\hat{L}_{0,1}-\log S(\theta-\theta_{1,1}).
    \end{aligned}
\end{equation}
The second TBA equation is singular at $\theta=\theta_{1,1}\pm \frac{\pi i}{2}$. The singularity at $\theta={\rm Re}(\theta_{1,1})$ corresponds to the zero point of $Y_{1,2}$.

Expanding the new TBA equation for small $\hbar$, we introduce
\begin{equation}
    -\log Y_{a,s}(\theta)\sim m_{s}e^{\theta}-m_{a,s}^{(\frac{1}{2})}+\sum_{n=1}^{\infty}m_{a,s}^{(n)}e^{(1-2n)\theta},
\end{equation}
where
\begin{equation}\label{eq:mcorr-K=1}
    \begin{aligned}
        m_{0,1}^{(n)}=&\frac{(-1)^{n}}{\pi}\int_{-\infty}^{\infty}e^{(2n-1)\theta}L_{1,2}(\theta)d\theta\\
        m_{1,2}^{(n)}=&\frac{(-1)^{n}}{\pi}\int_{-\infty}^{\infty}e^{(2n-1)\theta}\big(L_{0,1}+L_{2,1}\big)d\theta+(-1)^{n}\frac{2i}{2n-1}e^{(2n-1)\theta_{1,1}}.
    \end{aligned}
\end{equation}
We then test the new TBA equations numerically  by comparing the asymptotic expansions of the Y-function and the WKB periods at large $\theta$.  See Table \ref{tab:wkb-TBA-analytic}.

  \begin{table}[hpt]
    \centering
      \begin{tabular}
    {|c|c|c|c|c|}\hline
     $n$&    $m^{(n)}_{0,1}$ & $\Pi^{(n)}_{\gamma_1}/i$
     & $m_{1,2}^{(n)}$ & $\Pi_{\gamma_2}^{(n)}$\\\hline
        0&0.103932897990&0.103932897990&0.146983313914&0.146983313914\\
        1&0.344746764264&-0.344746764266&18.45478584239&18.454785842340\\
        2&67.13310204757&67.13310204761&-6.824849791937\ensuremath{\cdot10^{2}}&-6.824849791943\ensuremath{\cdot10^{2}}\\3&-5.994677109922\ensuremath{\cdot10^{4}}&5.994677110002\ensuremath{\cdot10^{4}}&1.020897181816\ensuremath{\cdot10^{6}}&1.020897181821\ensuremath{\cdot10^{6}}\\\hline
    \end{tabular}
    \caption{Comparison of the coefficients in the expansion based on the TBA equation
and the WKB periods at $m=0.6$ with $u_2=1/2$ and $E=1/64$. The TBA equations are solved numerically by using Fourier discretization with $2^{20}$ points and the cutoff $(-50,50)$. }
    \label{tab:wkb-TBA-analytic}
\end{table}

In this region of $m$, the  effective central charge is modified to
\begin{equation}
    \begin{aligned}
        \hat{c}_{\rm eff}(m)=&\frac{12}{\pi^{2}}\int d\theta\Big(m_{1}e^{\theta}\big(L_{0,1}+L_{2,1}\big)+m_{2}e^{\theta}L_{1,2}\Big)+i\frac{24}{\pi}m_{1}e^{\theta_{1,1}},
    \end{aligned}
\end{equation}
which can evaluated using the asymptotics of Y-function at $\theta\to -\infty$ \eqref{eq:const-sol}. The integration in $\hat{c}_{\rm eff}$ can be computed by using the Rogers dilogarithm function
\begin{equation}
    \begin{aligned}
       &\frac{12}{\pi^{2}}\int d\theta\Big(m_{1}e^{\theta}\big(L_{0,1}+L_{2,1}\big)+m_{2}e^{\theta}L_{1,2}\Big)\\
       =&\frac{12}{\pi^{2}}\Big({\cal L}_{1}\big(\frac{1}{1+Y_{1,2}^{\ast-1}}\big)+{\cal L}_{e^{2\pi im}}\big(\frac{e^{-2\pi im}}{1+Y_{0,1}^{\ast-1}}\big)+{\cal L}_{e^{-2\pi im}}\big(\frac{e^{2\pi im}}{1+Y_{2,1}^{\ast-1}}\big)\Big)\\&-i\frac{24}{\pi}(i\pi-m_{0,1}+m_{1}e^{\theta_{1,1}})-\frac{6}{\pi^{2}}\log(-1)L_{1,2}(-\infty),
    \end{aligned}
\end{equation}
which finally leads to
\begin{equation}
    \begin{aligned}
        \hat{c}_{\rm eff}(m)=4(1-8 m^2),\qquad\qquad \frac{1}{2}<m<\frac{3}{4}.
    \end{aligned}
\end{equation}
Express this in terms of the expansion of TBA equations, we find
\begin{equation}
    \hat{c}_{\rm eff}(m)=-\frac{12}{\pi}(m_{1}m_{2}^{(1)}+m_{2}m_{1}^{(1)}),\quad \frac{1}{2}<m<\frac{3}{4},
\end{equation}
where we used \eqref{eq:mcorr-K=1}. In terms of the WKB periods, one obtains the PNP relation \eqref{eq:PNP-Pi} once again.

\subsubsection{TBA for $3/4<m<1$}

The next singular point in TBA equations corresponds to the zeros, denoted by $\theta_{2,1}$, of $1+Y_{0,2}$. The TBA equations analytically continue to 
\begin{equation}
    \begin{aligned}
        \log\hat{Y}_{0,1}=&-m_{1}e^{\theta}+K\star L_{1,2}-\log S(\theta-\theta_{2,1}),\\
        \log Y_{1,2}=&-m_{2}e^{\theta}+K\star \hat{L}_{0,1}-\log S(\theta-\theta_{1,1}),
    \end{aligned}
\end{equation}
where $\theta_{1,1}$ and $\theta_{2,1}$ satisfy
\begin{equation}
    \begin{aligned}
        \log(-e^{-m_{0,1}})&=i\pi-m_{0,1}=-m_{1}e^{\theta_{_{1,1}}}+\int\frac{d\theta}{2\pi}\frac{L_{1,2}(\theta)}{\cosh(\theta_{1,1}-\theta)}-\log S(\theta_{1,1}-\theta_{2,1})
        ,\\
        \log(-1)=&i\pi=-m_{2}e^{\theta_{2,1}}+\int\frac{d\theta}{2\pi}\frac{\hat{L}_{0,1}(\theta)}{\cosh(\theta_{2,1}-\theta)}-\log S(\theta_{2,1}-\theta_{1,1}).
    \end{aligned}
\end{equation}

The effective central charge in this case is
\begin{equation}
    \begin{aligned}
        \hat{c}_{\rm eff}(m)=&\frac{12}{\pi^{2}}\int d\theta\Big(m_{1}e^{\theta}({L}_{0,1}+L_{2,1})+m_{2}e^{\theta}L_{1,2}\Big)+i\frac{24}{\pi}m_{1}e^{\theta_{1,1}}+i\frac{24}{\pi}m_{2}e^{\theta_{2}}\\
        =&4(1-8 m^2),\qquad \qquad \frac{3}{4}<m<1.
    \end{aligned}
\end{equation}

\subsection{Massless limit}\label{sec:massless}
We consider the WKB periods and the TBA equations in the limits where some one-cycles on the WKB curve degenerate. 
For the superpotential $W(x)=x^3/3-u_2/2 x$,
when the energy $E$ goes to zero, the allowed cycles $\gamma_1$ and $\gamma_3$ degenerate to a point. 
In the limit, the classical period $\Pi^{(0)}_{\gamma_1}$ goes to zero, while the period $\Pi^{(0)}_{\gamma_2}$ becomes a finite value of $2\sqrt{2} u_2^{3/2}/3$. 
When $E$ goes to $u_2^2/8$, $\gamma_2$ shrinks to zero and  $\Pi_{\gamma_1}^{(0)}\rightarrow 2i u_2^{3/2}/3$ and $\Pi_{\gamma_2}^{(0)}\rightarrow 0$.

Here we will consider the TBA equation in the $E\rightarrow 0$ limit.
The quantum corrections for the periods $\Pi_{\gamma_1}$ are finite in this limit and  are evaluated as
\begin{align}
\Pi^{(0)}_{\gamma_1}&=0,\\
\Pi^{(1)}_{\gamma_1}&=-2i\pi m,\\
\Pi^{(2)}_{\gamma_1}&={i(1-4m^2)\pi \over 4\sqrt{2}u_2^{3/2}},\\
\Pi^{(4)}_{\gamma_1}&={35 i(5-24m^2+16m^4)\pi\over 256\sqrt{2}u_2^{9/2}},\\
\Pi^{(6)}_{\gamma_1}&=-{3003 i (-53 + 268 m^2 - 240 m^4 + 64 m^6) \pi \over 
 16384 \sqrt{2} u_2^{15/2}}.
\end{align}
For $m=1/2$, we observe that $\Pi^{(n)}_{\gamma_{1}}=0$ ($n\geq 2$).
However, the corrections  $\Pi_{\gamma_2}^{(2n)}$ for the one cycle $\gamma_2$ are  divergent of order $O(1/E^{2n-1})$ in this limit.

In the region $|m|\leq {1\over2}$, where we have seen that the TBA equations 
\eqref{eq:TBA1} are valid,  let us take the limit $E\rightarrow 0$. Then the mass parameters have the limits  $m_1\rightarrow 0$ and $m_2\rightarrow 2\sqrt{2} u_2^{3/2}/3$ and the TBA equations \eqref{eq:TBA1} become
\begin{equation}\label{eq:massless-TBA1}
    \begin{aligned}
       \log Y_{0,1}=&2\pi i m+K\star L_{1,2},\\
       \log Y_{2,1}=&-2\pi i m+K\star L_{1,2},\\
       \log Y_{1,2}=&-m_{2}e^{\theta}+K\star\Big[L_{0,1}+L_{2,1}\Big].
    \end{aligned}
\end{equation}
Introducing a  new Y-function $\hat{Y}_{0,1}=e^{-2\pi i m}Y_{0,1}$, the TBA can be written as
\begin{equation}\label{eq:massless-TBA-new}
    \begin{aligned}
        \log\hat{Y}_{0,1}=&K\star L_{1,2}\\
        \log{Y}_{1,2}=&-m_{2}e^{\theta}+K\star\hat{L}_{0,1},
    \end{aligned}
\end{equation}
where $\hat{L}_{0,1}$ is defined in \eqref{eq:Lhat}. This system coincides with the conformal limit of the TBA equations of the sine-Gordon model at ${\cal N}=2$ supersymmetric point \cite{Fendley:1997ys}. 
The asymptotics  of the Y-functions are given by
\begin{align}
    \hat{Y}_{0,1}(\theta)&\rightarrow 2\cos{2\pi m\over3}, \quad Y_{1,2}(\theta)\rightarrow {\sin 2\pi m \over \sin{2\pi m\over3}}
\end{align}
for $\theta\rightarrow -\infty$, which is the same as the massive case. 
For $\theta\rightarrow +\infty$, the asymptotics are
\begin{align}
    \hat{Y}_{0,1}(\theta)&\sim 1 , \quad
    Y_{1,2}(\theta)\sim 2\cos{2\pi m\over2} e^{-m_2 e^\theta},
\end{align}
which is different from the massive case.
We can compute the Y-functions numerically and compare them with the quantum corrections to the WKB period. Table \ref{tab:masslessm01} shows that the massless limit of TBA equations describes the exact WKB period well. We also checked that at $m=1/2$, $m^{(n)}$ $(n=1,2,3)$ are of order $10^{-12}\sim 10^{-15}$, which is consistent with $\Pi^{(n)}_{\gamma_1}=0$. Although the coefficients in the asymptotic expansion of the WKB period are divergent, the Y-function is finite. It would be thus useful to take advantage of Y-function in the massless limit.
\begin{table}[]
    \centering
    \begin{tabular}{|c|c|c|c|c|}\hline
    $n$&    $m_{0,1}^{(n)}$ & $\Pi^{(n)}_{\gamma_1}/i$
    & $m_{1,2}^{(n)}$ & $\Pi_{\gamma_2}^{(n)}$\\\hline
    0&0&0&\ensuremath{1/3}&\ensuremath{1/3}\\1&-1.507964473727&1.507964473723&-1.904158772073\ensuremath{\cdot10^{21}}&\ensuremath{\infty}\\2&32.72282907987&32.72282907979&1.620951043969\ensuremath{\cdot10^{64}}&\ensuremath{\infty}\\3&-3.710585147572\ensuremath{\cdot10^{3}}&3.710585147575\ensuremath{\cdot10^{3}}&-2.573503923004\ensuremath{\cdot10^{107}}&\ensuremath{\infty}\\\hline
    \end{tabular}
    \caption{Comparison of the coefficients in the expansion based on the TBA equation
and the WKB periods at $m=0.1$ for $u_2=1/2$ and $E=0$. The TBA equations are solved numerically by using Fourier discretization with $2^{20}$ points and the cutoff $(-50,50)$.}
    \label{tab:masslessm01}
\end{table}

The original effective central charge \eqref{eq:ceff1} is the charge of the $D_3$ TBA equations. 
The constant solution of $\hat{Y}_{0,1}$ at $\theta\to \infty$ will also contribute to the calculation of the effective central charge \eqref{eq:ceff1}, which can be regarded as the effective central charge of $D_2$ TBA equations \cite{Ito:2019jio}. Subtracting this contribution, the effective central charge becomes
\begin{equation}
   \hat{c}_{\rm eff}(m)\Big|_{E=0}=\frac{12}{\pi^{2}}\int d\theta m_{2}e^{\theta}L_{1,2}=2\big(1-4m^{2}\big).
\end{equation}
As studied in \cite{Fendley:1997ys}, the TBA equations are analytically  continued at $m={1\over2},{3\over4},\dots$.
When $m>{1\over2}$, the singularity $\theta_{2,1}$ appears in the TBA equation, which becomes
\begin{align}\label{eq:tba_massless2}
    \log \hat{Y}_{0,1}&=K*L_{1,2},\\
    \log Y_{1,2}&=-m_2 e^{\theta}+K*\hat{L}_{0,1}-\log S(\theta-\theta_{1,1}).
\end{align}
Here $\theta_{1,1}$ is a solution of
\begin{align}
    -\pi i m&=-m_2 e^{\theta_{1,1}}+K*(L_{0,1}+L_{2,1}).
\end{align}
As $m$ increases from $1/2$ to $3/4$, ${\rm Re}( \theta_{1,1})$  changes from $-\infty$ to $\infty$. 
We compare the asymptotic expansion from the TBA equation \eqref{eq:tba_massless2}  with the WKB periods in Table \ref{tab:masslessm02}, which shows numerical agreement.
\begin{table}[]
    \centering
    \begin{tabular}{|c|c|c|c|c|}\hline
    $n$&     $m_{0,1}^{(n)}$ & $\Pi^{(n)}_{\gamma_1}/i$
    & $m_{1,2}^{(n)}$ & $\Pi_{\gamma_2}^{(n)}$\\\hline
    0&0&0&\ensuremath{1/3}&\ensuremath{1/3}\\1&1.507964473725&-1.507964473723&-1.112508703850\ensuremath{\cdot10^{22}}&\ensuremath{\infty}\\2&-20.05592750054&-20.05592750052&1.348108893562\ensuremath{\cdot10^{64}}&\ensuremath{\infty}\\3&2.080354914319\ensuremath{\cdot10^{3}}&-2.080354914323\ensuremath{\cdot10^{3}}&-2.322012143419\ensuremath{\cdot10^{107}}&\ensuremath{\infty}
    \\\hline
    \end{tabular}
    \caption{ Comparison of the coefficients in the expansion based on the TBA equation
and the WKB periods at $m=0.7$ for $u_2=1/2$ and $E=0$. The TBA equations are solved numerically by using Fourier discretization with $2^{20}$ points and the cutoff $(-50,50)$.}
    \label{tab:masslessm02}
\end{table}
At $m=3/4$, a new parameter $\theta_{2,1}$ appears in the TBA equation, and the full TBA for ${3\over4}<m<1$ is given by the massless limit of \eqref{eq:ac-TBA}. The modified effective central charge is
\begin{equation}
    \hat{c}_{\rm eff}(m)\Big|_{E=0}=\frac{12}{\pi^{2}}\int d\theta m_{2}e^{\theta}L_{1,2}+i\frac{24}{\pi}\sum_{j=1}^{J}m_{2}e^{\theta_{2,j}}.
\end{equation}
It is worth noting that the effective central charge $\hat{c}_{\rm eff}$ is related to the supersymmetric index $\log\Big({\rm Tr}\big[e^{i\alpha F}e^{-RH}\big]\Big)$ of two dimensional ${\cal N}=2$ theories \cite{Fendley:1997ys}, where $\alpha=2\pi m$ is the chemical potential of the fermion number. In particular, it becomes the Witten index (up to the overall factor) when $m=1/2$.

\section{The Voros spectrum}\label{sec:voros}
For the Schr\"odinger equation \eqref{eq:susy-sch}, starting from the turning points, we can draw the Stokes line, which will end on the other turning point or infinity. The Stokes lines divide the complex $x$-plane into several regions, called the Stokes regions. In each Stokes region, the Borel resummed WKB solution is continuous. Crossing the Stokes line, the basis of the WKB solution can have a jump, which is described by the connection formula \cite{Voros-83,Silverstone-85,AKT-91,DDP-97,DP-99,Iwaki:2014vad}\footnote{See also \cite{Sueishi:2020rug} for a nice review on this procedure.}. For the Schr\"odinger equation  \eqref{eq:susy-sch}, we can draw all the Stokes lines based on the un-deformed part $Q_0(x)$.
We  can find the WKB wave function in two regions using the connection formula. Imposing the boundary conditions on the wave function, one derives the exact quantization condition of the spectral problem which is expressed in terms of the exact WKB periods.
Since the exact quantization condition is fixed by the Stokes lines which does not depend on $Q_1(x)$, we thus expect the exact quantization condition for the case $W^\prime=x^2-1/4$ is the same as the one in the double well potential \cite{Zinn-Justin:1982hva,DDP-97}:
\begin{equation}
    \cos\Big(\frac{1}{\hbar}s_{{\rm med}}\big(\frac{1}{i}\Pi_{\gamma_{1}}(\hbar)\big)\Big)+\frac{1}{\sqrt{1+e^{-\frac{1}{\hbar}s(\Pi_{\gamma_{2}}(\hbar))}}}=0,
\end{equation}
which can be rewritten as the condition
\begin{equation}
    \frac{1}{\hbar}s_{{\rm med}}\big(\frac{1}{i}\Pi_{\gamma_{1}}(\hbar)\big)+\epsilon \arctan\Big(e^{-\frac{1}{2\hbar}s(\Pi_{\gamma_{2}}(\hbar))}\Big)=2\pi (k+\frac{1}{2}),\quad k\in \mathbb{Z}_{\geq 0}
\end{equation}
with  parity parameter $\epsilon=\pm 1$.
Note that each $k$ leads to two spectra depending on the parity $\epsilon$. The level of the spectrum can thus be labelled from the bottom by
\begin{equation}
    k_\epsilon=2k-\frac{\epsilon-1}{2}=0,1,\cdots.
\end{equation}
In terms of the Y-functions, the WKB periods can be expressed by using
\begin{equation}
    \begin{aligned}
        \frac{1}{\hbar}s_{{\rm med}}\big(\frac{1}{i}\Pi_{\gamma_{1}}\big)(\hbar)&=\frac{i}{2}\big(\log Y_{0,1}(\theta+\frac{\pi i}{2}+i0)+\log Y_{0,1}(\theta+\frac{\pi i}{2}-i0)\big)\\
        &=m_{1}e^{\theta}+im_{0,1}^{(\frac{1}{2})}+\mbox{ P}\int\frac{d\theta^{\prime}}{2\pi}\frac{\log\big(1+Y_{1,2}(\theta^{\prime})\big)}{\sinh(\theta-\theta^{\prime})},\\
        \frac{1}{\hbar}s(\Pi_{\gamma_{2}})(\hbar)=&-\log Y_{1,2}(\theta).
    \end{aligned}
\end{equation}
Here ${\rm P}$ denotes the principal value, which can be computed by
\begin{equation}
    {\rm P}\int\frac{f(\theta^\prime)}{\sinh(\theta-\theta^\prime)}d\theta^\prime=\lim_{\delta\to 0}\int\frac{\sinh(\theta-\theta^\prime)\cos(\delta)f(\theta^\prime)}{\sinh^2(\theta-\theta^\prime)\cos^2(\delta)+\cosh^2(\theta-\theta^\prime)\sin^2(\delta)}d\theta^\prime
\end{equation}

To solve the exact quantization condition of the form ${\cal Q}(\hbar, E,\{u\})=0$, we will usually fix the Plank constant $\hbar$ and the moduli parameter $\{u\}$ of the potential, and  we solve the condition about the energy $E$. One can also reverse this procedure by fixing the energy $E$ and moduli parameter $\{u\}$ at first, and then one solves the condition about  $\hbar$ \cite{Voros-83}. Those two procedures describe the same curve in the  $(E,\hbar)$-space.

\begin{figure}
    \centering
    \begin{tabular}{cc}
    \resizebox{70mm}{!}{\includegraphics{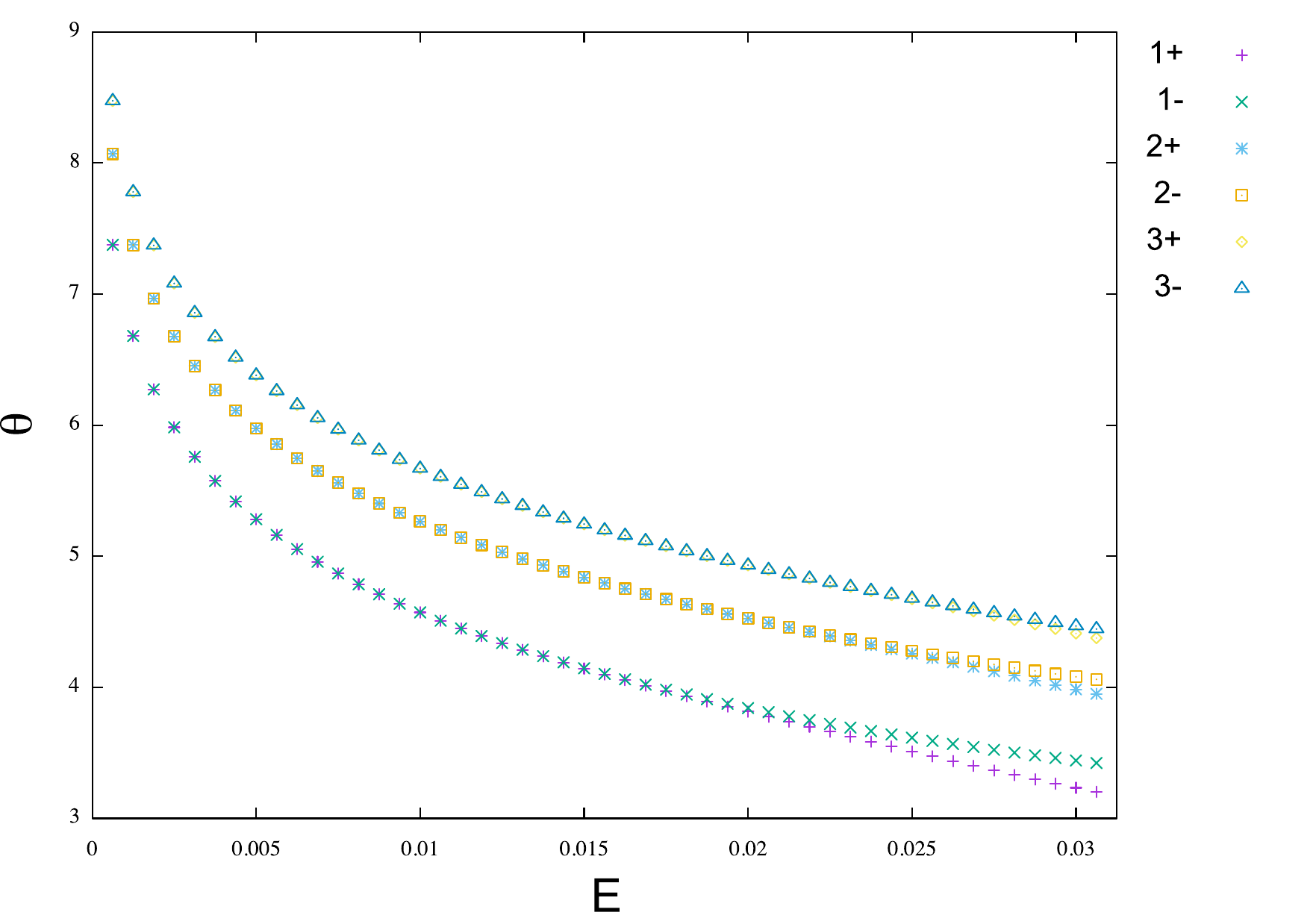}} &
    \resizebox{70mm}{!}{\includegraphics{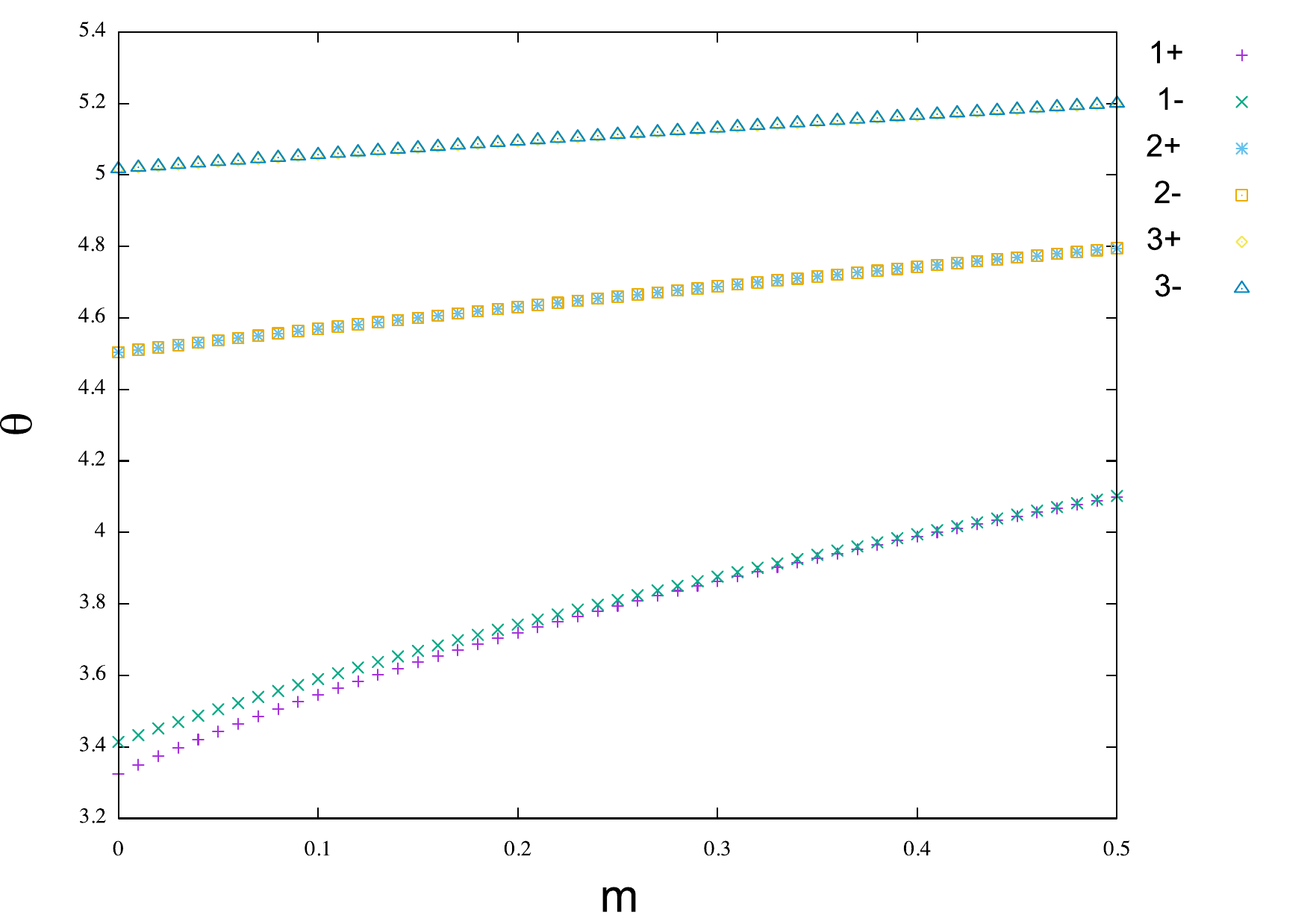}}\\
    (a) & (b)\\
    \end{tabular}
    \caption{(a) The Voros spectrum for $0\leq E\leq 1/32$ at $m=1/2$ and (b) for $0\leq m\leq 1/2$ at $E=1/64$. The TBA equations are solved numerically by using Fourier discretization with $2^{20}$ points and the cutoff $(-50,50)$.}
    \label{fig:voros_m_theta}
\end{figure}

We will solve the exact quantization condition about $\hbar$ at level $k_\epsilon$ via the TBA equations, which leads to the Voros spectrum $\theta_{k_\epsilon}=-\log \hbar_{k_\epsilon}$. On the other hand, we can regard the Voros spectrum $\theta_{\ast,n}$ as input and then compute the energy spectrum.
In the table \ref{tab:vorosm12}, the Voros spectrum for $u_2=1/2$ and  $E=1/64$ are calculated from the exact quantization condition.
The corresponding energy spectrum is recalculated by the diagonalization of the Hamiltonian using the basis of the eigenfunction of the harmonic oscillator \cite{https://doi.org/10.1002/qua.26554}. It is found that the eigenvalues agree with the original value.
In Figure \ref{fig:voros_m_theta}, we show the $E$ and $m$-dependence of the Voros spectrum. In Fig. \ref{fig:voros_m_theta} (a), the $1_{\pm}$ spectra degenerate for small $E$ but split when $E$ becomes large due to the tunnelling effect between two local minima.

\begin{table}[]
    \centering
    \begin{tabular}{|c|c|c|}
\hline
$k_{\epsilon}$ & $\theta_{k_\epsilon}$ & $E_{k_{\epsilon}}$ \\
\hline
\ensuremath{1_{+}}&4.098461939440&0.015625000025\\\ensuremath{1_{-}}&4.101928506979&0.0156250000105\\\ensuremath{2_{+}}&4.794626227596&0.015625000012\\\ensuremath{2_{-}}&4.794647371948&0.015625000011\\
\ensuremath{3_{+}}&5.200323939648&0.015625000010\\\ensuremath{3_{-}}&5.200323939648&0.015625000010 \\
\hline
\end{tabular}
    \caption{Voros spectrum for $u_2=1/2$, $E=1/64=0.015625$, $m=1/2$ and the corresponding energy obtained from the diagonalization of the Hamiltonian with 200 eigenfunctions of the harmonic oscillator. The TBA equations are solved numerically by using Fourier discretization with $2^{20}$ points and the cutoff $(-50,50)$.}
    \label{tab:vorosm12}
\end{table}

\section{Conclusions and Discussion}\label{sec:con}

In this paper, we have studied the ODE/IM correspondence for the Schr\"odinger equation of  the deformed supersymmetric quantum mechanics including the $\hbar$-linear deformation. The TBA equations describing the exact WKB periods are obtained from the ${\mathbb Z}_4$ extension of the undeformed TBA equations. For the Schr\"odinger equation with $\hbar$ deformed double well potential, the TBA equations reduce to two sets of D-type TBA equations. Varying the deformation parameter $m$ in the deformed term of the potential, a series of poles appear in the integral of the TBA equations. Subtracting the poles, one obtains the analytic continuation of the TBA equations which matches with the results of the exact WKB analysis. We also studied the zero mass limit of the TBA equations corresponding to the ground state energy of the model, which is found to agree with the WKB periods. 
Moreover, combining the exact quantization condition, our TBA equations are shown to reproduce the energy spectrum. 

There are many open questions raised by the present work. Let us list some of them. In this paper, we have broken the parity symmetry of the double-well potential by adding a $\hbar$ deformed term in the potential. In the meantime, the $A_3$-type TBA system corresponding to the double-well potential equations becomes a TBA system with two decoupled D-type TBA. It would be interesting to consider the deformation in the quantum integrable model side such that these two TBA systems are connected.

In this paper, we have focused on the minimal chamber of the moduli space of the potential. Increasing the energy above the barrier of the potential, a wall-crossing of the TBA equations occurs where the cycles for the WKB periods change. In the present case, it is  related to the wall-crossing of the D-type TBA equations derived in \cite{Ito:2019jio}. It would be interesting to study the wall-crossing in detail for the $\hbar$ deformed potential. We can also apply our method to the Schr\"odinger equation with  undeformed monomial potential and a $\hbar$ deformed term, whose TBA equations should be consistent with the ones obtained from the wall-crossing. 

In section \ref{sec:massless}, we have considered the case where some one-cycles become degenerate. This case appears for the ground state of the supersymmetric quantum mechanics with $\hbar$ deformed double-well potential, whose energy 
is known to be zero. Since the Stokes graph will be modified to the degenerate Weber type from the Airy type, the resurgence structure will be modified as well \cite{Kamata:2021jrs}. It would be thus important to study the correspondence between TBA and WKB to describe the resurgence structure for the Schr\"odinger equation with degenerated turning points \cite{DDP-97}.

It is also interesting to investigate analytic continuations and wall-crossing phenomena of the TBA equations for the quantum SW curves for ${\cal N}=2$ SQCD and related superconformal field theories, whose ODE includes the  $\hbar$-corrections in the potential,   for  the study of strong coupling physics.

\subsection*{Acknowledgements}
We would like to thank Daniele Gregori, Syo Kamata, Yong Li, Marco Rossi, JingJing Yang and  Hao Zou for useful discussions.
The work of K.I. is supported in part by Grant-in-Aid for Scientific Research 21K03570 from Japan Society for the Promotion of Science (JSPS).
The work of H.S. is supported by the start-up Funding of Zhengzhou University. H.S. would like to thank Tokyo Institute of Technology for their hospitality.

\begingroup\raggedright\endgroup

\end{document}